\documentstyle{mn}

\def\spose#1{\hbox to 0pt{#1\hss}} 
\def\simlt{\mathrel{\spose{\lower 3pt\hbox{$\mathchar"218$}}
     \raise 2.0pt\hbox{$\mathchar"13C$}}}
\def\simgt{\mathrel{\spose{\lower 3pt\hbox{$\mathchar"218$}}
     \raise 2.0pt\hbox{$\mathchar"13E$}}}

\begin{document}

\title[The supersoft source RX~J0527.8-6954]
 {Optical identification of the LMC supersoft source RX~J0527.8-6954 from 
MACHO Project photometry}

\author[C. Alcock et al.]
{C. Alcock$^{1,2}$, R.A. Allsman$^{3}$, D.R. Alves$^{1,4}$, 
T.S. Axelrod$^{1,5}$, D.P. Bennett$^{1,2}$,  \and P. A. Charles$^{6}$, 
K.H. Cook$^{1,2}$, K.C. Freeman$^{5}$, K. Griest$^{2,7}$, 
M.J. Lehner$^{2,7}$, \and M. Livio$^{8}$,  
S.L. Marshall$^{2,9}$, D. Minniti$^{1}$, 
B.A. Peterson$^{5}$, M.R. Pratt$^{2,9}$, \and
P.J. Quinn$^{5}$, A.W. Rodgers$^{5}$, 
K.A. Southwell$^{6}$, C.W. Stubbs$^{2,9,10}$, \and 
W. Sutherland$^{6}$ and D.L. Welch$^{11}$\\
$^1$ Lawrence Livermore National Laboratory, Livermore, CA 94550, USA\\
$^2$ Center for Particle Astrophysics, University of California, Berkeley, 
CA 94720, USA\\
$^3$ Supercomputing Facility, Australian National University, Canberra, A.C.T. 
0200, Australia\\
$^4$ Department of Physics, University of California, Davis, CA 95616\\ 
$^5$ Mt.  Stromlo and Siding Spring Observatories, 
Australian National University, Weston, A.C.T. 2611, Australia\\
$^6$ University of Oxford, Department of Astrophysics, Nuclear \& Astrophysics
Laboratory, Keble Road Oxford, OX1 3RH\\
$^7$ Department of Physics, University of California,
San Diego, CA 92093, USA\\
$^8$ Space Telescope Science Institute, 3700 San Martin Drive, Baltimore, 
MD~21218, USA\\
$^9$ Department of Physics, University of California, Santa Barbara, CA 93106, 
USA\\
$^{10}$ Departments of Astronomy and Physics, University of Washington, Seattle, 
WA 98195, USA\\
$^{11}$ Department of Physics and Astronomy, Mc~Master University, Hamilton, 
Ontario, Canada, L8S~4M1\\}

\date{Received
      in original form	}

\maketitle

\begin{abstract}

We identify the likely optical counterpart to the LMC supersoft X-ray source 
RX~J0527.8-6954, and hence recover HV\,2554. 
This identification is based on an analysis of $\sim 4$~years 
of optical photometry obtained serendipitously via the MACHO project. We see 
a steady fading of the star of $\sim 0.5$~mag over the duration of the 
observations. Evidence is also presented for an orbital modulation of 
$\sim 0.05$~mag semi-amplitude on a period of $P=0.39262\pm0.00015$~d. 
Our optical observations are consistent with the suggestion that the X-ray 
decline in this system is caused by cooling after a weak shell flash. 

\end{abstract}


\section{Introduction}

The supersoft sources (SSS) are a group of X-ray objects characterised by their extremely luminous ($L_{\rm bol} \sim~10^{38}$~erg\,s$^{-1}$) 
emission at energies $<0.5$~keV. The prototypical sources, CAL~83 and CAL~87, 
discovered by the {\it Einstein} X-ray observatory (Long, Helfand \& Grabelsky 
1981), are accreting binaries which share optical similarities with the low 
mass X-ray binaries (e.~g.\ Pakull et al.\ 1988; Cowley et al.\ 1993). 
Initially, the nature of the compact object remained elusive, with 
suggestions of both black hole (Cowley et al.\ 1990) and neutron star 
accretors 
existing (Greiner, Hasinger \& Kahabka 1991; Kylafis \& Xilouris 1993). 

However, the model which has gained predominance involves 
white dwarf primaries accreting at sufficiently high rates ($\dot{M} \simgt 
10^{-7} {\rm M}_{\odot}$~yr$^{-1}$) to sustain steady nuclear surface burning 
(van den Heuvel et al.\ 1992). Thus, the extremely soft X-ray emission at 
near-Eddington limited luminosities is naturally explained. In order to 
achieve these rates, the donor star is required to be more massive than the 
white dwarf, so that thermally unstable mass transfer can occur. This 
model has recently received observational support from optical spectroscopy 
of the system RX~J0513.9-6951, which exhibits collimated outflows at typical 
white dwarf escape velocities (Southwell et al.\ 1996). 

{\it ROSAT} observations have considerably enlarged the group of supersoft 
X-ray sources (e.~g.\ Kahabka \& Tr\"{u}mper 1996). The object 
RX~J0527.8-6954, discovered in the {\it ROSAT} first-light observations 
(Tr\"{u}mper et al.\ 1991), is a transient SSS, having been undetected in 
earlier {\it Einstein} observations. A blackbody  
fit to the {\it ROSAT} spectrum (Greiner, Hasinger \& Kahabka 1991) 
indicated spectral parameters very similar to CAL~83. However, 
continued {\it ROSAT} monitoring of the source has revealed an  
exponential decline in the X-ray luminosity by a factor of $\sim 50$ 
from 1990-1994 (Greiner et al.\ 1996), a behaviour currently unique 
among the SSS to this object. 

The identification of the optical counterpart has proved extremely 
elusive (e.~g.\ Cowley et al.\ 1993). 
Greiner (1996) presents a detailed investigation of a search for 
the Harvard variable star 
HV~2554, which lies within the X-ray error circle, and is thus a prime 
candidate. However, he concludes that the variability of this star is 
questionable, and that no obvious optical signatures are present 
in any of the resolvable stars within the X-ray error circle (see 
Fig.~2 of Greiner 1996). 

Given the relatively long timescale of the X-ray variability ($\sim$~yrs), it 
is clearly advantageous to obtain extended optical monitoring over a 
similar timebase - serendipitously, the MACHO project affords us such an 
opportunity. Hence, we are able to identify the probable optical 
counterpart (presumably HV~2554) by looking for long term optical trends in 
the stars in the X-ray error box, over a period of $\simgt 4$~yrs.

\section{Observations}

The MACHO project (Alcock et al.\ 1995a) involves nightly monitoring of 
certain LMC/SMC fields for microlensing events. The LMC supersoft source 
RX~J0527.8-6954 is located in one of these fields, hence we have an optical 
history of this source since the beginning of the project in August 1992.  
The observations were made using the 
$1.27$-m telescope at Mount Stromlo Observatory, Australia. 
A dichroic beamsplitter and filters provide simultaneous CCD photometry 
in two passbands, a `red' band ($\sim 6300-7600$~\AA) 
and a `blue' band ($\sim 4500-6300$~\AA); the latter is 
approximately equivalent to the Johnson $V$ filter. 

The images were reduced with the standard MACHO photometry code {\sc
SoDoPHOT}, based on point-spread function fitting and differential
photometry relative to bright neighbouring stars. Further details of
the instrumental set-up and data processing may be found in Alcock et
al.\ (1995b), Marshall et al.\ (1994) and Stubbs et al.\ (1993).

\section{Results}

We show in Fig.~1 the template image used by the profile fitting routine to 
determine the resolvable stellar objects in the vicinity of RX~J0527.8-6954. 
Stars which were resolved, and therefore for which there exist individual 
light curves, are numbered according to the notation of Greiner (1996). We 
therefore have photometry for all the objects listed 1-9, which lie within 
the X-ray error circle. 

An inspection of the light curves of all these objects revealed interesting 
variability in Star~6. We should caution that, given the crowded nature of 
the field and variable seeing conditions, this could be contaminated by light 
from Star~9. Our definition of ``the counterpart'' is therefore to be taken as 
Star~6 or Star~9 - higher resolution time-resolved studies will be required 
to resolve this ambiguity. However, we note that Greiner et al.\ (1996) also 
commented on the apparent variability of Star~6 when compared with the image 
of Cowley et al.\ (1993). 

\subsection{Light curve}

We show in Fig.~2 the red and blue photometry of the object flagged as
Star~6. The data consists of MACHO project observations taken during
the period 1992 August~8 -- 1996 November~2. The absolute calibration of the
MACHO fields and transformation to standard passbands is not yet
complete, thus the measurements are plotted differentially relative to
the observed median.  One-sigma error bars are shown. 
A linear fit to each dataset reveals a steady decline of 0.12~mag/yr 
in the blue and 0.085~mag/yr in the red. Over the time-span of the dataset, 
which is 4.25~yrs, the optical fading thus amounts to 0.51~mag in the blue and 
0.36~mag in the red.

\subsection{Period Analysis and Folded Light Curve}

The long term decline was removed with a linear fit before performing a power 
spectrum analysis on the red and blue datasets. 
We used a Lomb-Scargle technique (Lomb 1976; Scargle 1982) and show the 
resulting periodograms in Fig.~3. A frequency space of 
$0.05-10$~cycles~d$^{-1}$ was searched with a resolution of 
0.001~cycles~d$^{-1}$. There is a dominant 
peak in both datasets at $P=0.39262\pm0.00015$~d, with lesser power at 
0.6477~d and 0.28193~d (and 1.838~d in the blue data only). 
We checked the significance of all the 
peaks by analysing the power spectra of randomly 
generated datasets, using the sampling intervals of the real data. 
Our Monte Carlo simulations reveal that the 0.39262~d period occurs with at 
least 4-$\sigma$ confidence in both datasets. In the red data, the remaining 
peaks occur with only $\sim 1-\sigma$ confidence or less. In the blue data, 
all four principle peaks represent at least 3-$\sigma$ detections; however, 
assuming the stongest peak is the true orbital period (0.39262~d), 
we note that the 0.6477~d and 0.28193~d periods occur at the values we 
would expect for one-day aliases, having frequencies which differ by 
1 cycle~d$^{-1}$ from the dominant peak. Furthermore, the 1.838~d period, 
(which appears only in the blue data) has a frequency consistent with a 
two-day alias. 

We therefore find strong evidence to suggest an orbital period of 
$P=0.39262\pm0.00015$~d. 
The red and blue detrended data were filtered to reject any data points with 
errors exceeding 0.2~mag. We then folded the remaining points on a 
period of 0.39262~d to examine the form of any orbital modulation. 
We find that the blue data are well fitted by a 
sinusoid of semi-amplitude 0.052$\pm$0.009 mag, suggesting a low 
inclination. This is shown in Fig.~4. 
We derive an ephemeris of T$_{\rm o}={\rm JD}\,244 8843.07(1) + 0.3926(2)E$, 
where T$_{\rm o}$ is the time of maximum optical brightness, and $E$ is an 
integer. 

The fit to the red folded data was less good and is not shown. This is 
probably due to the larger number of anomalous points in the red light 
curve (day number $\sim 1300-1500$ in Fig.~2). 
We have no convincing explanation for these points, there being no 
indications of particularly bad seeing or instrumental effects which might 
have accounted for them. 

\subsection{Long term and orbital colour variation}

We investigated the relative MACHO B-R colour as a function of time. Our 
data indicate a steady reddening of the system, amounting to a total of 
$\sim 0.15$~mag over the 1546 days of observations. By removing this long-term 
trend with 
a linear fit, and folding the data on $P=0.39262$~d, we found evidence for a 
sinusoidal modulation in the B-R colour of semi-amplitude 
$\sim 0.02$~mags. These folded data also indicated the system 
to be reddest at $\phi \sim 0.5$, namely the time of minimum light.

\section{Discussion}

\subsection{Simultaneous optical/X-ray behaviour}

During the period 8/8/1992 - 10/8/1995, we may 
investigate the simultaneous X-ray behaviour using the {\it ROSAT} data 
presented in Greiner et al.\ (1996). The optical decline (in the blue 
MACHO observations) during this time was 0.36~mags (a factor of $\sim 1.4$), 
which may be compared with a decrease by a factor of $\sim 5.9$ in the X-rays. 
The system also became redder by $\sim 0.1$~mag in this time. 

Greiner et al.\ (1996) suggested that the decline in the X-rays represents 
a decrease in the temperature, following perhaps a weak shell flash. The 
observed decline in the optical is consistent with such a picture, since
some of the optical luminosity is probably produced by the irradiation
of the secondary star and the accretion disc by the hot white dwarf. A small
decline in the mass transfer rate may also be expected due to the decreased
irradiation of the secondary (e.g. Livio 1992). We can obtain a rough estimate
of the decay time by examining the Kelvin-Helmholtz timescale of the envelope: 
\begin{equation}
\tau_{\rm{\small KH}} \approx \frac{GM_{\rm{\small WD}} \Delta m_{\rm env}}
{R_{\rm{\small WD}} L_{\rm{\small WD}}},
\end{equation}
where $L_{\rm{\small WD}}$ is the luminosity and $\Delta m_{\rm env}$ is the 
mass of the envelope. If we approximate the envelope mass by that required
to produce a thermonuclear runaway on a cold WD (which should be regarded
as an upper limit; Yungelson et al.\ 1995), then we obtain for the decay time: 
(scaled with the parameter values of a $1 M_{\odot}$ white dwarf), 
\begin{eqnarray}
\tau_{\rm decay} \simlt \tau_{\rm KH}  \approx 2300~{\rm days} 
\left(\frac{M_{\rm{\small WD}}}{1 M_{\odot}} \right)^{0.2} 
\left(\frac{R_{\rm{\small WD}}}{7.8 \times 10^{-3} R_{\odot}} \right)^{2.2}  
\left(\frac{L_{\rm{\small WD}}}{10^{38}{\rm ~erg~s}^{-1}}\right)^{-1}. 
\end{eqnarray} 
Given the observed decay time in the X-rays of $\simgt 5.5$~years (Greiner et 
al. \ 1996), 
this suggests that the WD has a mass of $\sim 1 M_{\odot}$. 
An examination of the results of Prialnik \& Kovetz (1995) also suggests
a mass of about $1 M_{\odot}$, if the decay time represents the time for 
cessation of nuclear burning.

\subsection{Binary parameters}

Assuming an orbital period of $P=0.39262$~d, we may calculate the 
mean density, $\overline{\rho}$, of the companion star under the 
assumption that it fills its Roche lobe. We combine Kepler's 
third law and 
the Eggleton (1983) relation for a Roche-lobe filling star: 
\begin{equation}
\frac{R_{L_2}}{a}=\frac{0.49q^{-2/3}}{0.6q^{-2/3} + 
\mbox{ln}(1+q^{-1/3})}, 
\end{equation}
where $R_{L_2}$ is the radius of a sphere with the same volume as the 
secondary Roche lobe, $a$ is the binary separation, and $q$ is the binary 
mass ratio ($\equiv {\rm M}_{\scriptsize{compact}}/
{\rm M}_{\scriptsize{secondary}}$) to obtain:
\begin{equation}
\overline{\rho} = \frac{0.161}{{\rm P}^2(1+q)} \left 
(0.6 + q^{2/3} \ln(1+q^{-1/3}) \right )^3 {\rm g}~{\rm cm}^{-3}, 
\end{equation} 
where $P$ is in days. 
For values of $q \simlt 1$, as 
required by the van den Heuvel et al.\ (1992) model, the implied mean 
density is $\sim 1.1$~g\,cm$^{-3}$. This is consistent with that of a 
$\sim 1.3~{\rm M}_{\odot}$ main sequence star of spectral type $\sim$ 
F5 (Allen 1973), although, for the obtained 
orbital period, the secondary star could be slightly evolved.  
We estimated the mass 
of the compact object to be $\sim 1 M_{\odot}$. 
Hence, values of $q \simlt 1$, 
as required by the van den Heuvel et al.\ (1992) model, are indeed possible 
if the orbital period assumed here is correct.

\section{Summary}

We identify the optical counterpart of the supersoft source 
RX~J0527.8-6954 from MACHO Project 
photometry. A steady fading of $\simlt 0.5$~mags in the optical light 
is observed over $\sim 4$~years. 
We detect an orbital modulation of 
$\sim 0.05$~mag semi-amplitude on a period of $P=0.39262\pm0.00015$~d. 

We find the optical fading to be consistent with the model for the X-ray 
decline suggested by Greiner et al.\ (1996), in which the system is 
cooling after a weak shell flash. We estimated the decay time using the 
Kelvin-Helmholtz timescale of the accreted hydrogen envelope, and obtain a 
likely white dwarf mass of $\sim 1 M_{\odot}$. Combining this with the 
inferred mass of the secondary yields a mass ratio which is consistent with 
the van den Heuvel et al.\ (1992) model for the supersoft X-ray sources. 

\newpage
\subsection*{Acknowledgments}

We are grateful for the support given our project by the technical
staff at the Mt. Stromlo Observatory. Work performed at LLNL is 
supported by the DOE under contract W-7405-ENG. Work performed by the
Center for Particle Astrophysics personnel is supported by the NSF 
through grant AST 9120005. The work at MSSSO is supported by the Australian
Department of Industry, Science and Technology. 
KG acknowledges support from DoE OJI, Alfred P. Sloan, and Cotrell Scholar 
awards. 
CWS acknowledges the generous support of the Packard and Sloan Foundations.
WS and KAS are both supported by PPARC through an Advanced Fellowship and 
studentship respectively. 
The data analysis was performed using the Starlink {\sc period} package at the 
University of Oxford Starlink node.

\newpage

\begin{figure}
\caption{MACHO template image of the field around RX~J0527.8-6954, showing 
stars identified for profile fitting. We resolve the objects listed 1-8 
considered by Greiner (1996), and additionally Star~9. 
North is up, East to the left.} 
\end{figure}

\begin{figure}
\caption{The optical light curve of RX~J0527.8-6954 from MACHO project
observations. The relative magnitude is shown for the red and `blue'
filters (the latter is approximately equivalent to the Johnson $V$
passband). Note the overall decline in both light curves of $\sim  
0.5$~mag (blue) and $\sim 0.4$~mag (red) in $\simgt 4$~years.}
\end{figure}

\begin{figure} 
\caption{The Lomb-Scargle periodograms of RX~J0513-69 MACHO time 
series data in the blue filter ({\it upper}) and the red filter ({\it lower}). 
The strongest peak is at $P=0.39262\pm0.00015$~d in both datasets.}
\end{figure}
 
\begin{figure}
\caption{The blue light curve of RX~J0527.8-6954, folded on a period of 
0.39262~d. The data have been averaged into 45 phase bins, and 
are fitted with a sinusoid of amplitude 0.052~mag.}
\end{figure}

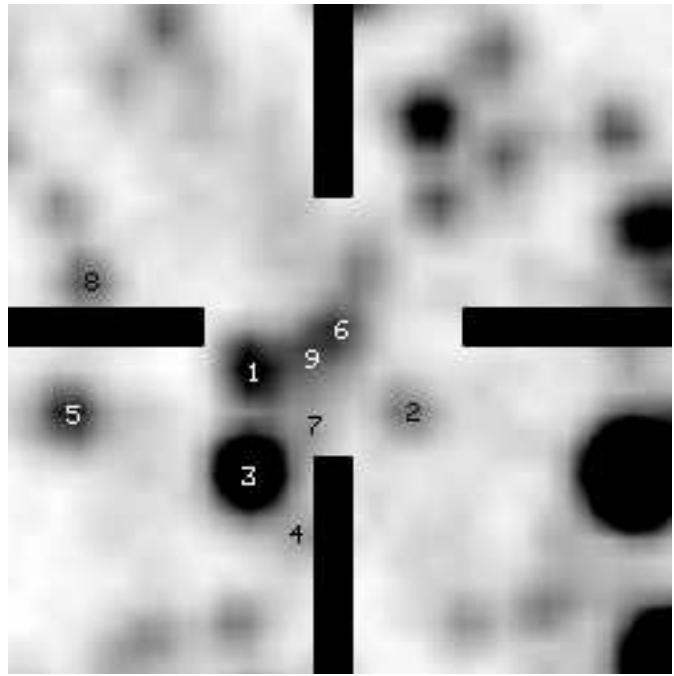
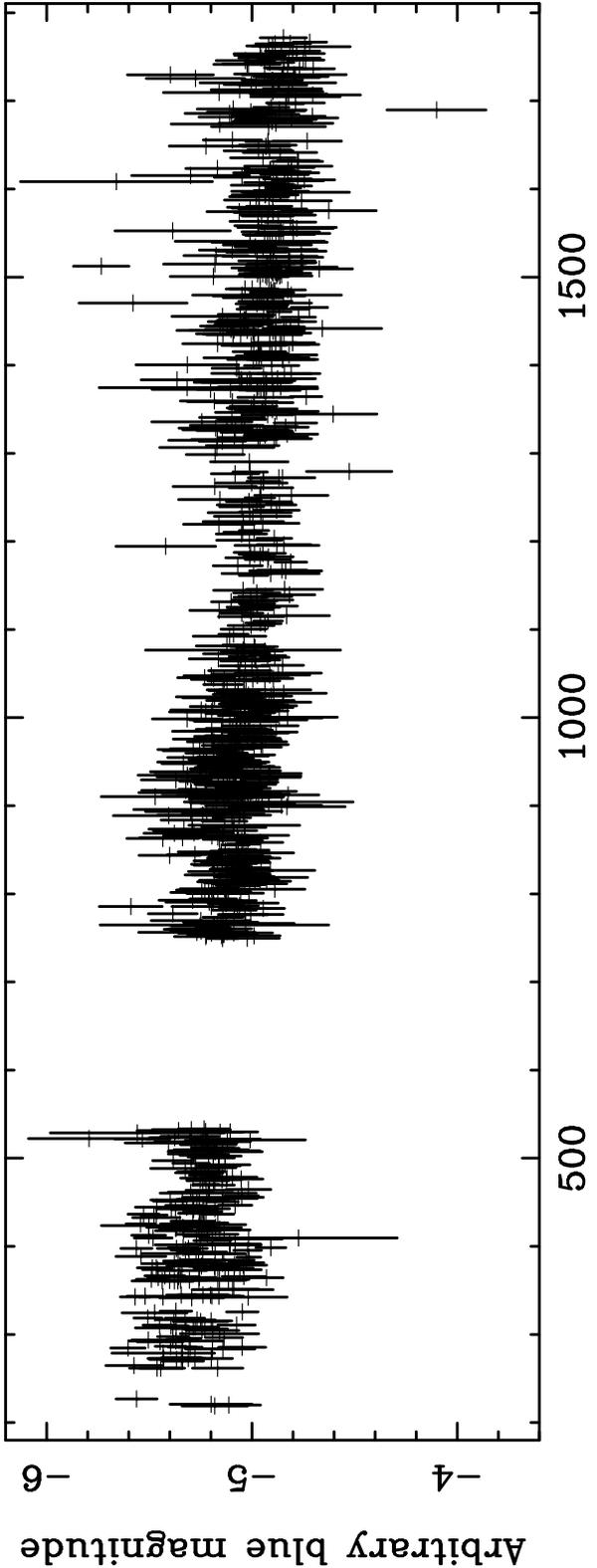
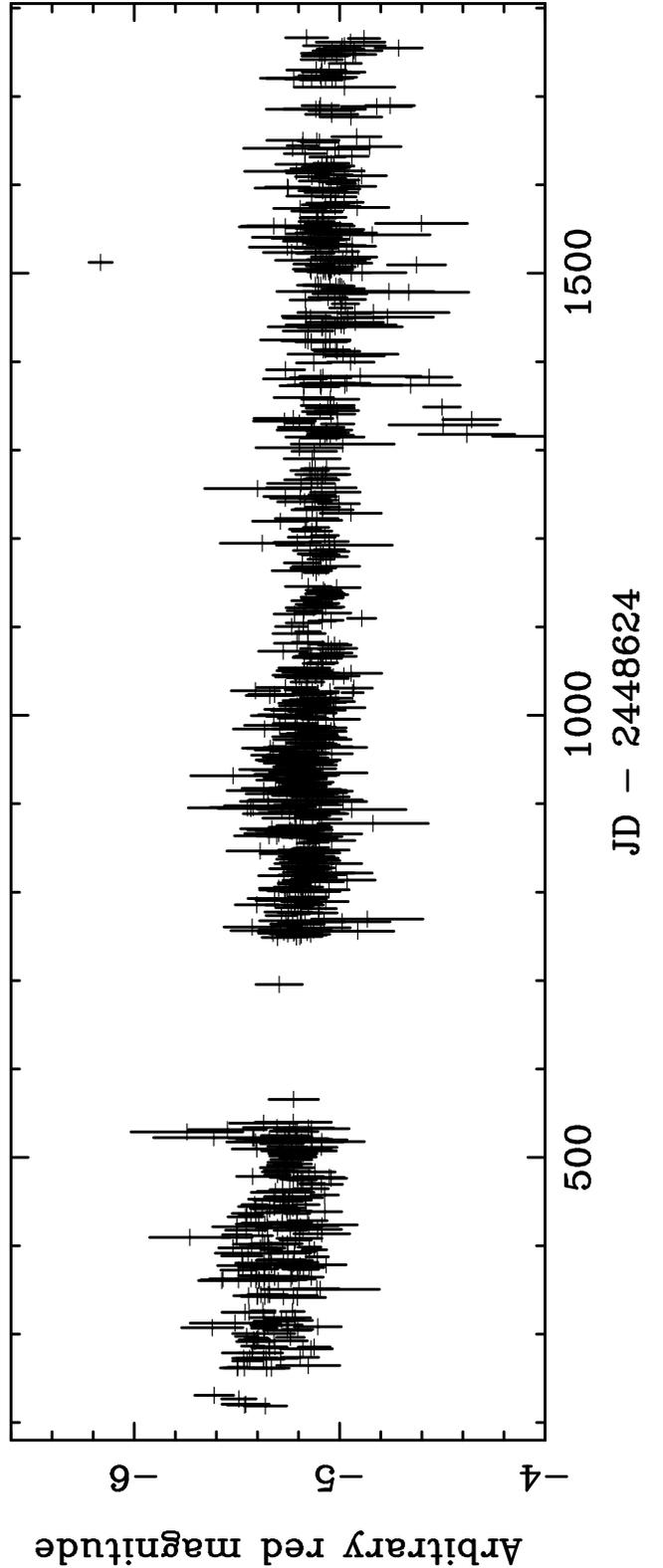
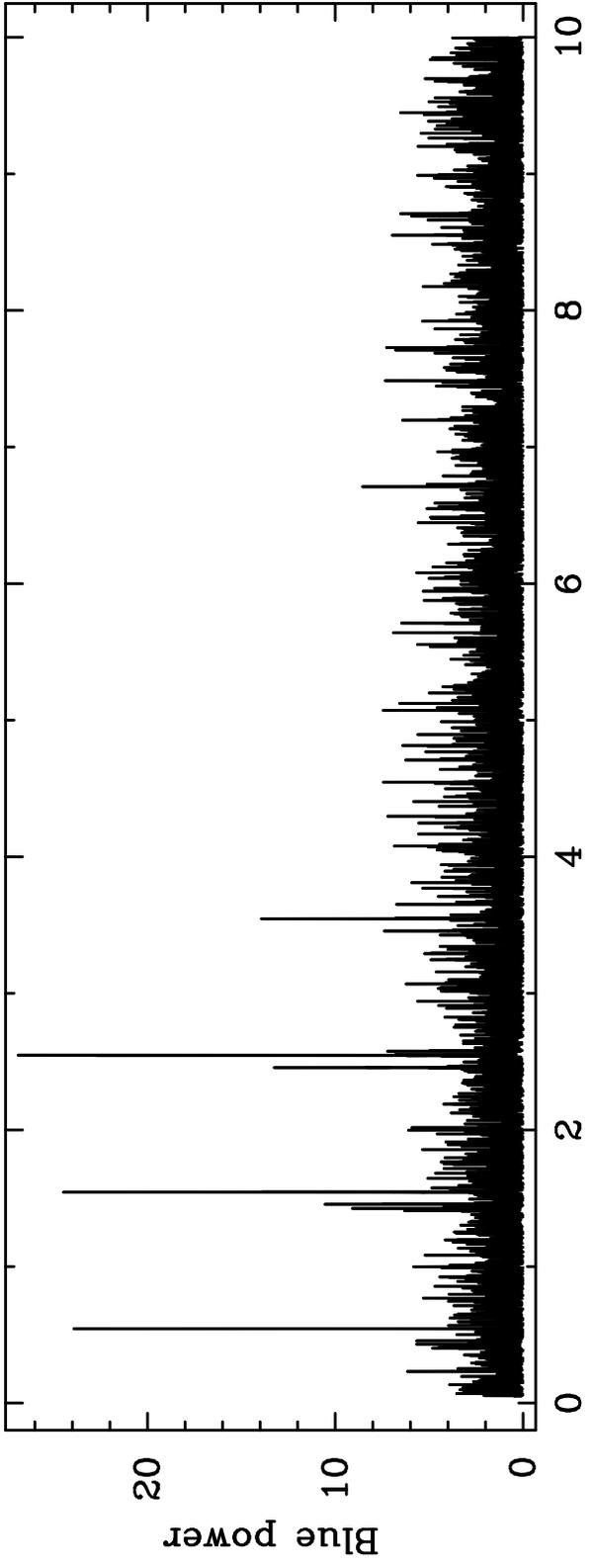
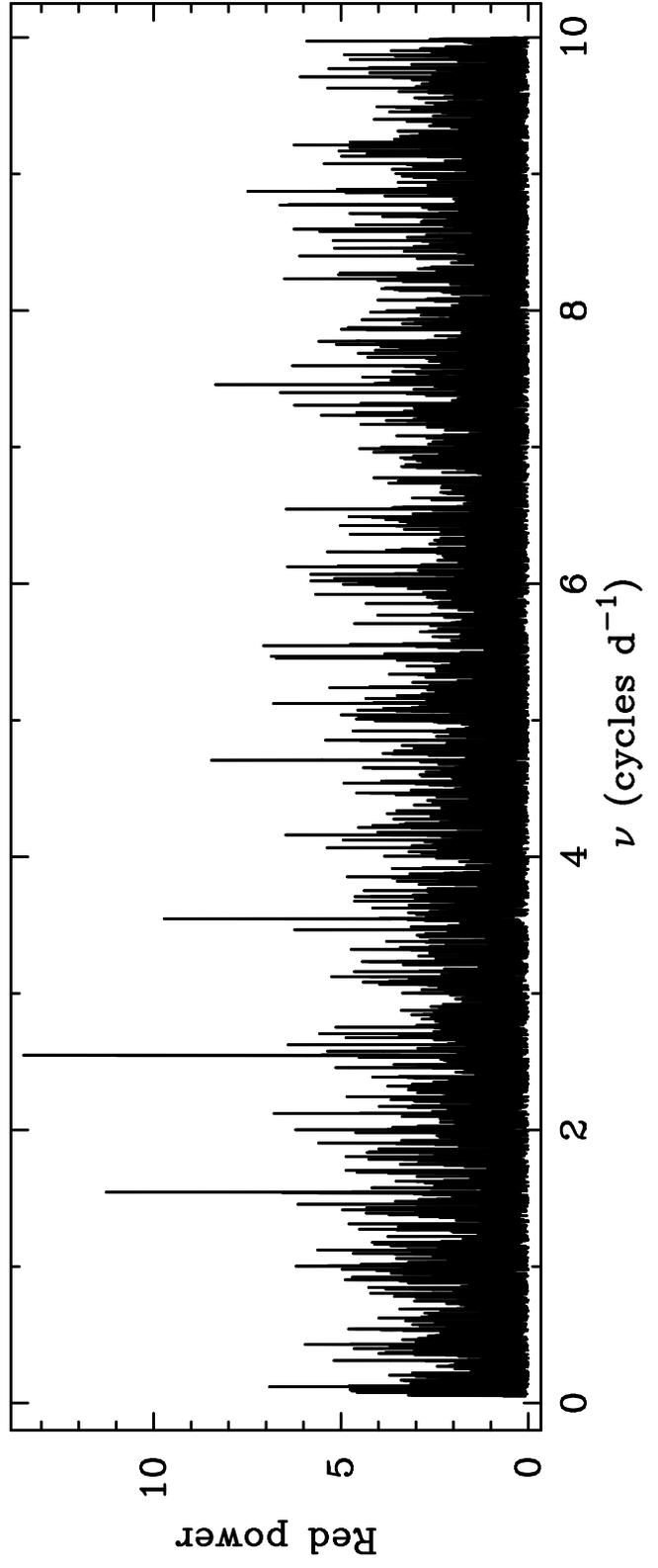
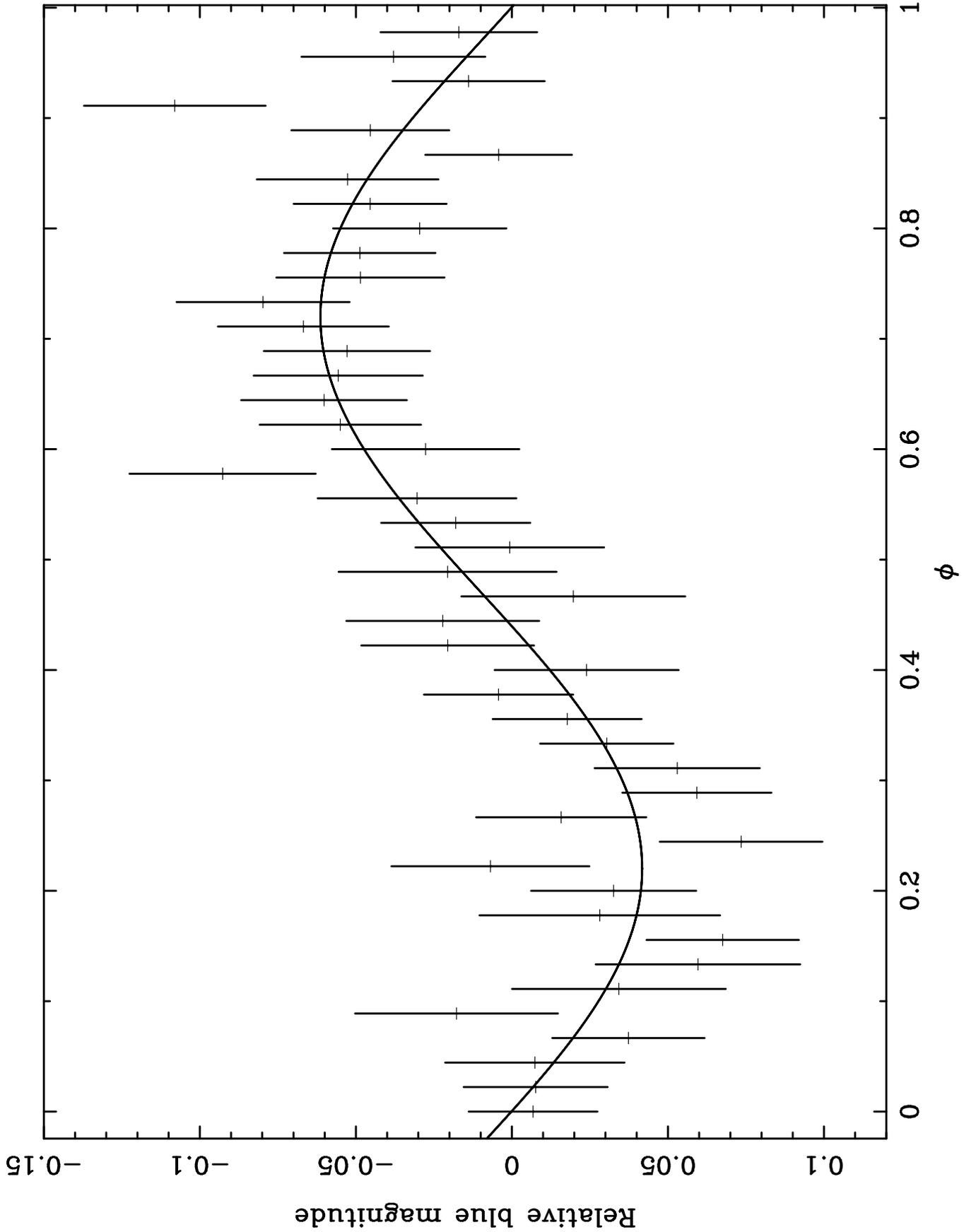
\end{document}